\documentclass[conference]{IEEEtran}

\usepackage{graphicx}
\usepackage{bm}
\usepackage{enumerate}
\usepackage{float}
\usepackage{multicol}
\usepackage{color}
\usepackage{url}
\usepackage{cite}
\usepackage{flushend}

\usepackage{amsmath}
\usepackage{amsthm}
\usepackage{amssymb}
\DeclareMathOperator*{\argmax}{argmax}

\usepackage{algorithmic}
\usepackage{algorithm}
\usepackage{acronym}
\usepackage{array}

\ifCLASSOPTIONcompsoc
  \usepackage[caption=false,font=normalsize,labelfont=sf,textfont=sf]{subfig}
\else
  \usepackage[caption=false,font=footnotesize]{subfig}
\fi

\acrodef{BS}{base station}
\acrodef{RA}{receive antenna}
\acrodef{PA}{predictor antenna}
\acrodef{IID}{independent and identically distributed}
\acrodef{CDF}{cumulative distribution function}
\acrodef{PDF}{probability density function}
\acrodef{cu}{channel use}
\acrodef{ACK}{acknowledgement}
\acrodef{NACK}{negative acknowledgement}
\acrodef{SNR}{signal-to-noise ratio}
\acrodef{HARQ}{hybrid automatic repeat request}
\acrodef{CSIT}{channel state information at the transmitter side}
\acrodef{INR}{incremental redundancy}
\acrodef{npcu}{nats-per-channel-use}
\acrodef{MIMO}{multiple input multiple output }
\acrodef{TDD}{time division duplex}
\acrodef{E2E}{end-to-end}
\acrodef{IAB}{integrated access and backhaul}
\acrodef{3GPP}{3rd Generation Partnership Project}
\acrodef{MR}{moving relay}
\acrodef{QoS}{quality of service}
\acrodef{VPL}{vehicle penetration loss}
\acrodef{LoS}{line-of-sight}
\acrodef{NLoS}{non-line-of-sight}
\acrodef{mmw}{millimeter wave}
\acrodef{CoMP}{Coordinated Multi-Point}
\acrodef{MISO}{multiple input single output }
\acrodef{BF}{beamforming}
\acrodef{FBL}{finite block-length}
\acrodef{UE}{user equipment}

\begin{document}
\captionsetup{belowskip=0pt,aboveskip=0pt}

\title{Velocity-aware Antenna Selection in \\Predictor Antenna Systems}
\author{\IEEEauthorblockN{Hao Guo}
\IEEEauthorblockA{\textit{Department of Electrical Engineering} \\
\textit{Chalmers University of Technology}\\
Gothenburg, Sweden \\
hao.guo@chalmers.se}
\and
\IEEEauthorblockN{ Behrooz Makki}
\IEEEauthorblockA{\textit{Department of Ericsson Research} \\
\textit{Ericsson Research}\\
Gothenburg, Sweden\\
behrooz.makki@ericsson.com}
\and
\IEEEauthorblockN{ Tommy Svensson}
\IEEEauthorblockA{\textit{Department of Electrical Engineering} \\
\textit{Chalmers University of Technology}\\
Gothenburg, Sweden\\
tommy.svensson@chalmers.se}
}

% \author{Hao~Guo,
%         Behrooz~Makki,~\IEEEmembership{Senior~Member,~IEEE},
%         and Tommy~Svensson,~\IEEEmembership{Senior~Member,~IEEE}% <-this % stops a space
% \thanks{H. Guo and T. Svensson are with the Department of Electrical Engineering, Chalmers University of Technology, 41296 Gothenburg, Sweden (email: hao.guo@chalmers.se; tommy.svensson@chalmers.se).}% <-this % stops a space
% \thanks{B. Makki is with Ericsson Research, 41756 Gothenburg, Sweden (email: behrooz.makki@ericsson.com).}}% <-this % stops a space
%\thanks{M.-S. Alouini is with the King Abdullah University of Science and Technology,
%Thuwal 23955-6900, Saudi Arabia (e-mail: %slim.alouini@kaust.edu.sa).}}
%\thanks{Manuscript received April 19, 2005; revised January 11, 2007.}}

\maketitle

\begin{abstract}
Moving relay (MR), which is a candidate solution for supporting in-vehicle users, has been investigated in different studies. Due to the mobile nature of the MR, acquiring channel state information at the transmitter side (CSIT) is challenging because of the fast-changing environment around the vehicle. On  top of an MR, one can use predictor antenna (PA), i.e., an additional antenna in front of the receive antenna (RA), to obtain CSIT, and recent works have investigated the benefits of such a set up. PA-aided CSIT acquisition normally works with the help of different content information such as the location and the velocity of the MR. In this paper, we study the effect of velocity awareness on the PA system, and develop adaptive antenna selection schemes in PA-assisted MRs. Results show that, compared to no-CSIT schemes, a velocity-aware antenna selection-based PA system can improve the end-to-end throughput by an order of magnitude.
\end{abstract}


\section{Introduction}
As a complement to fixed relays, \ac{MR} has been discovered to be a candidate method to serve mobile users \cite{yutao2013moving,guo2021predictor}. \acp{MR} can improve the \ac{QoS} of in-vehicle users by reducing the \ac{VPL} and performing group handover.  One main challenge for \ac{MR} is, however, how to obtain accurate \ac{CSIT} because the relay itself is naturally moving. To overcome this issue, the concept of \ac{PA} has been first proposed in \cite{Sternad2012WCNCWusing} in which two sets of antennas are deployed on top of a vehicle.  The \ac{PA} positioned in the front is used for predicting the \ac{CSIT} for the antennas behind it, referred to as \ac{RA}, who will later encounter a similar channel as the \ac{PA} encountered several time slots before. 

Various testbed-based studies have validated the \ac{PA} concept together with advanced technologies such as massive \ac{MIMO}  \cite{phan2018WSAadaptive} and low-pass filtering \cite{BJ2017ICCWusing} at high velocity. It is shown in \cite{BJ2017PIMRCpredictor} that the \ac{PA} system can be beneficial in both \ac{TDD} and FDD (F: Frequency) setups with \ac{LoS} and NLoS (N: non) channels. Moreover, using  Kalman filtering to interpolate for the \ac{TDD} uplink estimation is studied in \cite{Apelfrojd2018PIMRCkalman}.

The \ac{PA} system, however, may face spatial mismatch, i.e., the \ac{RA} does not reach the same spatial point as the \ac{PA}, due to varying vehicle velocity or the processing delay at the \ac{BS} \cite{guo2021predictor,guo2020semilinear}. That is, for the acquired \ac{CSIT} to be accurate, the \ac{RA} needs to be as close as possible to the point where the \ac{PA} has estimated the channel. Such an alignment scheme, and/or alternative methods (e.g., rate adaptation in \cite{guo2020semilinear}) require knowledge of vehicle velocity, control loop time, as well as antenna spacing. \cite{guo2020semilinear} focuses on SISO (S: Single) setup and proposes analytical models of the \ac{PA} system with spatial mismatch, while the effect of multiple antennas at both transceiver sides has not been fully investigated.

In this paper, we first review the \ac{PA} concept and give a comprehensive literature study of the \ac{PA}. Then, we study the data transmission efficiency and reliability of \ac{PA} systems with multiple antennas at the \acp{MR}. Specifically, we investigate the effect of knowing the velocity on the system performance. Also, we design an adaptive scheme to choose the best \ac{RA} based on the vehicle velocity, such that spatial mismatch is minimized. Finally, we study the effect of different parameters such as the antennas distance and the vehicle velocity on the performance of \ac{PA} systems.  The results show that, using adaptive antenna selection and rate adaptation, there is potential for improving the performance of the \ac{PA} system.

\section{PA Concept and Literature Review}
Thanks to the flexible deployment and the \ac{VPL} elimination, \ac{MR} has the potential to become to one important component in mobile networks \cite{yutao2013moving,guo2021predictor,shan2021CST}. The benefit of deploying \ac{MR} can be mainly divided into two aspects:
\begin{itemize}
    \item{\textit{Eliminating \ac{VPL} for in-vehicle users}}: The users inside the vehicles such as commuters  normally expect similar, uninterrupted mobile service as they have in home or office. However, vehicle itself forms a natural blockage to the radio signal propagation which leads to high \ac{VPL} \cite{tanghe2008TVT}. Thus, \ac{VPL} can drastically degrade the signal strength and as a result, affect the received \ac{QoS} remarkably. Results in \cite{tanghe2008TVT} reveal that the \ac{VPL} is around 25 dB at sub-6 GHz and it could definitely be worse at higher frequencies such as \ac{mmw} bands, due to the fact that \ac{mmw} signals are more sensitive to propagation blockers. Using \ac{MR} with both outside-vehicle and inside-vehicle modules, the \ac{VPL} can almost be eliminated and the users inside the vehicle could experience a nearly pure \ac{LoS} condition \cite{yutao2013moving}.
    \item{\textit{Increasing the coverage of cellular networks for out-vehicle users}}: As one type of relay with relatively easy deployment, \ac{MR} can also be used for extending the coverage of macro-cell network by cooperative relying \cite{li2012tvt,khan2017tvt}. By joint transmission with macro-cell \acp{BS}, the \ac{MR}-aided system can remarkably improve the coverage probability \cite{li2012tvt}. Indeed, compared to serving the in-vehicle \acp{UE}, data transmission to out-vehicle \acp{UE} is of lower priority, unless for, e.g., public safety use-cases where the \ac{MR} is used for coverage extension.
\end{itemize}

In both applications of the \ac{MR}, outside-vehicle backhaul links have attracted more attention from researchers than inside-vehicle access links since the later one is normally \ac{LoS} with low design complexity. Also, wireless backhaul is more suitable to deploy in outdoor \acp{MR} because of its flexibility and low cost. Naturally, the design of the wireless backhaul link becomes the bottleneck of the system capacity in order to serve many \ac{MR}-assisted users \cite{yutao2013moving}. 

Using advanced transmission schemes at the \ac{BS} side such as \ac{MIMO}, \ac{CoMP} joint transmission, and different rate allocation algorithms can improve the performance of the \ac{MR} backhaul link. Among all these techniques \ac{CSIT} plays an important role since the quality of \ac{CSIT} affects the system performance directly. On the other hand, with moving vehicle as \ac{MR}, the environment around changes quickly and the channel estimation delay makes the \ac{CSIT} at the \ac{BS} side outdated quickly. The situation becomes more severe at higher moving speed and frequency bands. 

One can use diversity-based transmission schemes at the cost of extra resources. Also, using location information can help with \ac{CSIT} estimation but the location uncertainty needs to be handled carefully \cite{srikar2016twc}. Alternatively, using channel predictors such as Kalman or Wiener can obtain the prediction range of 0.1-0.3 times the wavelength, which is sufficient for low/moderate speed vehicles or pedestrians  at cm-wave carrier frequencies \cite{Sternad2012WCNCWusing}. Nevertheless, these predictors become inefficient when speed and frequency increases.  Compensate by increasing the number of pilots and perform channel interpolation would introduce unrealistic overhead.

As one promising setup aiming at acquiring better \ac{CSIT} condition for \acp{MR}, the \ac{PA} setup was firstly proposed in 
2012 \cite{Sternad2012WCNCWusing}, where two groups of antennas are mounted on the top of a vehicle. The front one w.r.t. moving direction is dedicated as \ac{PA}(s) which is(are) used for channel sensing and estimation and the obtained channel information is sent back to the \ac{BS}. Then, in the next time slot(s), the vehicle moves forward and the antennas behind \ac{PA}, denoted as \acp{RA}, would finally reach the same spatial point as the \acp{PA}. In this way, if the processing delay at the \ac{BS} is designed properly based on, e.g., wavelength, antenna separations and vehicle speeds, the \acp{RA} could receive the signal at the point where the \ac{PA} used to estimate the channel. By doing so the \ac{CSIT} for the \acp{RA} can achieve high accuracy and as a result, the system is capable of utilizing various advanced transmission schemes to improve the \ac{QoS}.

In \cite{Sternad2012WCNCWusing}, compared with Kalman filter-based prediction, it is shown that a \ac{PA}-assisted system can give decent \ac{CSIT} accuracy in both \ac{LoS} and \ac{NLoS} conditions with prediction range being around $\lambda$ and $\lambda/2$, respectively, with $\lambda$ being the wavelength. Following \cite{Sternad2012WCNCWusing}, different testbed- and simulation-based studies have investigated and revealed the potential of the \ac{PA}. \cite{yutao2013moving} introduces dedicated \ac{MR} links with deployment of the \ac{PA}, and studies the effect of \ac{VPL} by system-level simulations. Using multiple antennas at the \ac{BS}, \cite{Dinh2013ICCVEadaptive} design a downlink \ac{MISO} system with \ac{PA} interpolation, in order to mitigate residual \ac{BF} mis-pointing. From antenna design perspective, \cite{Jamaly2014EuCAPanalysis} focus on the \ac{PA}-\ac{RA} correction and performance coupling compensation to enhance the performance. Using antenna coupling compensation, at least three times $\lambda$ can be predicted with accuracy above 96\% cross-correlation between the \ac{PA} and \ac{RA} channels \cite{Jamaly2014EuCAPanalysis}. With early development of 5G, \cite{PhanHuy2016GLOBECOM5g} investigates optimal required number of \acp{PA}/\acp{RA} on the top of a vehicle in terms of spectrum efficiency and power, at both 1.2 and 3.5 GHz. 

Besides modeling and simulations, testbed-based research studies have been performed to validate the feasibility of the \ac{PA} concept. Part of \cite{Jamaly2014EuCAPanalysis} is dedicated to the demonstration of the proposed coupling compensation in downtown Dresden, Germany. Results in \cite{Jamaly2014EuCAPanalysis}  indicate the superior performance of using quarter-wavelength monopoles rather than half-wavelength dipoles in the \ac{PA} system. Then, in 2017, a testbed study was performed at 2.53 GHz in Dresden, verifying that the prediction horizon of the \ac{PA} system can be at least up to three times the wavelength \cite{BJ2017ICCWusing}. As an extension of \cite{BJ2017ICCWusing}, \cite{BJ2017PIMRCpredictor} proposes interpolation-based \ac{PA} estimation scheme and verifies it with practical measurements. Finally, with massive \ac{MIMO}, \cite{phan2018WSAadaptive} evaluates the performance of the \ac{PA} system at 2.18 GHz, and the large gains on the  signal-to-interference ratio is a promising indication of the feasibility of utilizing the  \ac{PA} in multiple antenna systems.

Given that 1) the \ac{PA} system is highly sensitive to spatial mismatch, i.e., the \acp{RA} could not reach the same point as the \acp{PA}; 2) The \ac{PA} system does not have analytical model to obtain tractable evaluations, recent studies \cite{Guo2019WCLrate,guo2020semilinear,guo2020power,Guo2020rate,guo2020comnet,guo2021predictor,guo2021mag2} address these problem and further clarify the potential of the \ac{PA} system. \cite{Guo2019WCLrate} investigates the spatial mismatch problem in the \ac{PA} system, and proposes a tractable channel model to evaluate the effect of spatial variation with rate adaptation. Then, the model in \cite{Guo2019WCLrate} has been further evaluated in \cite{guo2020semilinear} with temporal evolution of the channel. With two sets of antennas on  top of a vehicle, the potential of involving \ac{PA} partially into the transmission have been studied in \cite{guo2020power,Guo2020rate}, where \ac{HARQ} is combined with system and optimal power \cite{guo2020power} and average rate \cite{Guo2020rate} of the \ac{HARQ}-based \ac{PA} system are investigated respectively. Considering finite block-length transmission, \cite{guo2020comnet} studies the effect of the codeword length on the average throughput and error probability. Also, \cite{guo2021predictor} reviews the recent progress on the \ac{PA} study and proposes adaptive/non-adaptive delay methods to be applied in the \ac{PA} system with spatial mismatch. The role of the \ac{PA} in the development of \ac{MR} and \ac{IAB} are also summarised. Finally, \cite{guo2021mag2} extends the \ac{PA} concept to multiple vehicles and investigates the potential of using \ac{PA} for dynamic blockage avoidance in cooperative internet-of-vehicle networks.

In \cite{Guo2019WCLrate,guo2020semilinear,guo2020power,Guo2020rate,guo2020comnet,guo2021predictor,guo2021mag2}, the speed of vehicle is assumed to be known at the \ac{BS}, which may not be practical in real applications. Alternatively, using the speed information at the vehicle side and perform antenna selection could potentially enhance the \ac{PA} system at the presence of spatial mismatch.

\section{Model and Method}
\begin{figure}
\centering
  \includegraphics[width=1.0\columnwidth]{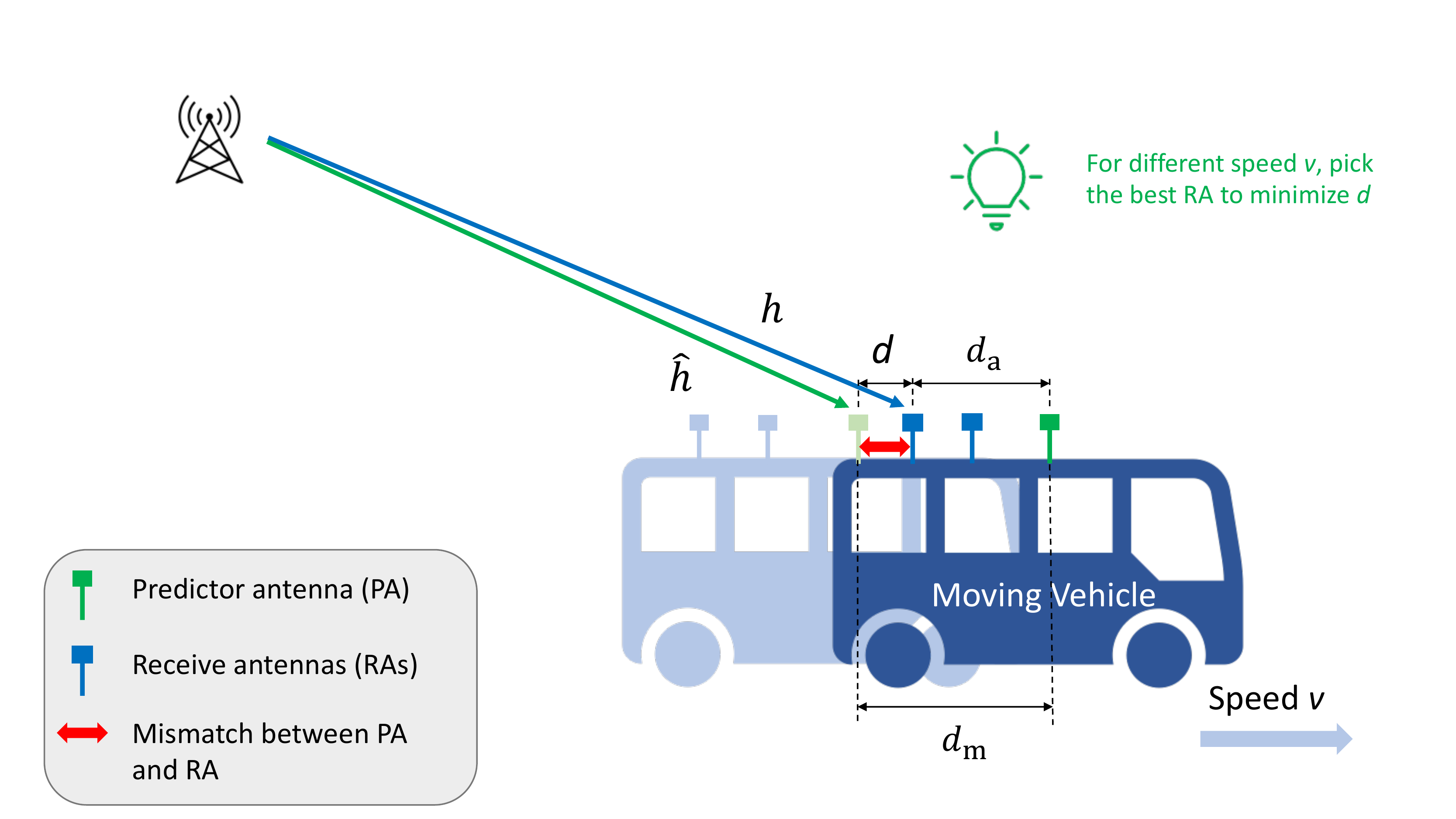}\\
\caption{The illustration of the \ac{PA} system with spatial mismatch.}
\label{fig_1}
\end{figure}
Considering a setup illustrated in Fig. \ref{fig_1}, a moving vehicle is deployed with one \ac{PA} and multiple \acp{RA} on its top. The separation between the PA and $i$-th \ac{RA} is $d_{\text{a},i}$. First, let's consider the case with one \ac{RA} to introduce the model, i.e., the separation of the \ac{PA} and the \ac{RA} is $d_{\text{a}}$, as used in, e.g., \cite{guo2020semilinear}. At time $t_1$, the \ac{PA} estimates the channel and sends the channel information back to the \ac{BS}. Then, at $t_2$, depending on, e.g., $d_\text{a}$, speed $v$, and the processing time $t_{\text{BS}}$ at the \ac{BS}, the \acp{RA} may end up in different places. In this way, the mismatch distance $d$ between 
\begin{itemize}
    \item the place where the \ac{PA} estimates the channel, and
    \item the point where the \ac{RA} actually reaches at $t_2$
\end{itemize}
can be expressed by
\begin{align}\label{eq_d}
    d = |vT - d_{\text{a}}|,
\end{align}
where $T = t_2 - t_1$. Denote $y$ as the received signal at the \ac{RA}, and it is given by
\begin{align}\label{eq_Y}
y = \sqrt{p}hx + z.
\end{align}
Here, $P$ is transmit power, and $x$ represents transmit signal with unit energy. $z \sim \mathcal{CN}(0,1)$ is the independent and identically distributed complex Gaussian noise added at the receiver side. Using the same method as in \cite{guo2020semilinear} 
\begin{align}\label{eq_H}
    h = \sqrt{1-\sigma^2} \hat{h} + \sigma q
\end{align}
to model the \ac{BS}-\ac{RA} channel $h$ as a function of the \ac{BS}-\ac{PA} channel $\hat{h}$ and $q \sim \mathcal{CN}(0,1)$. Here, $\sigma$ is a parameter which is a function of $d$ in (\ref{eq_d}), and larger $\sigma$ represents bigger spatial mismatch.  Defining $\hat{g} \doteq (1-\sigma^2)|\hat{h}|^2$ with $h \sim \mathcal{CN}(0,1)$ and $ g \doteq |{h}|^2$, the CDF of $g$ is given by
\begin{align}\label{eq_cdf}
    F_{g|\hat{g}}(x) = 1 - Q\left( \sqrt{\frac{2\hat{g}}{\sigma^2}}, \sqrt{\frac{2x}{\sigma^2}}  \right).
\end{align}
Here, $Q(,)$ represents Marcum-Q function. 

\begin{algorithm}[tbp!]
 \caption{Velocity-aware antenna selection scheme in PA systems with spatial mismatch}
 \begin{algorithmic}
%  \REQUIRE in
%  \ENSURE  out
\STATE \textit{In each time slot with known velocity at the receiver side, do the followings:} 
\begin{enumerate}[I.]
\item Calculate mismatch distance $d_i$ for each \ac{RA} using (\ref{eq_d}), with known information of $v$, $T$ and $d_{\text{a},i}$.
\item Evaluate the expected throughput (\ref{eq_etaexpected}) using proper methods for each \ac{RA}.
\item Pick the best \ac{RA} with the highest expected throughput.
  \end{enumerate}
 \end{algorithmic}
 \label{alg_1}
 \end{algorithm}

The goal is to maximize the outage/error-limited throughput which is defined as 
\begin{align}\label{eq_etaexpected}
  \eta = \mathbb{E}\left\{\eta|{\hat {g}}(r_{|\hat{g}}^{\text{opt}})\right\},
\end{align}
with
\begin{align}\label{eq_optR}
    r_{|\hat{g}}^{\text{opt}}&=\argmax_{r_{|\hat{g}}\geq 0} \left\{ \left(1-\Pr\left(\log(1+gP)<r_{|\hat{g}}\right)\right)r_{|\hat{g}} \right\}.
    \end{align}

In \cite{Guo2019WCLrate,guo2020semilinear}, different analytical methods of solving (\ref{eq_optR}) are presented, and results indicate that while rate adaptation can mitigate spatial mismatch and provide notable throughput gain compared to no \ac{CSIT}, the throughput is still very sensitive to the speed variation with one \ac{RA}. 

Given that the speed is known at the receiver side, it is possible to add additional \acp{RA} and perform antenna selection to obtain better performance. The proposed velocity-aware antenna selection scheme is summarized in Algorithm \ref{alg_1}. Here, we assume that in each time slot the velocity is known at the receiver side. In this way, from (\ref{eq_d}) the mismatch distance for each \ac{RA} $d_i$ can be calculated. Then, we evaluate the expected throughput (\ref{eq_etaexpected}) for each \ac{RA}. Finally, the best \ac{RA} can be picked with the highest throughput, which can be expressed as
\begin{align}\label{eta_best}
    \eta^{\text{opt}} = \max_{i} \eta_i,
\end{align}
where $\eta$ is given by (\ref{eq_etaexpected}).

\begin{figure}
\centering
  \includegraphics[width=1.0\columnwidth]{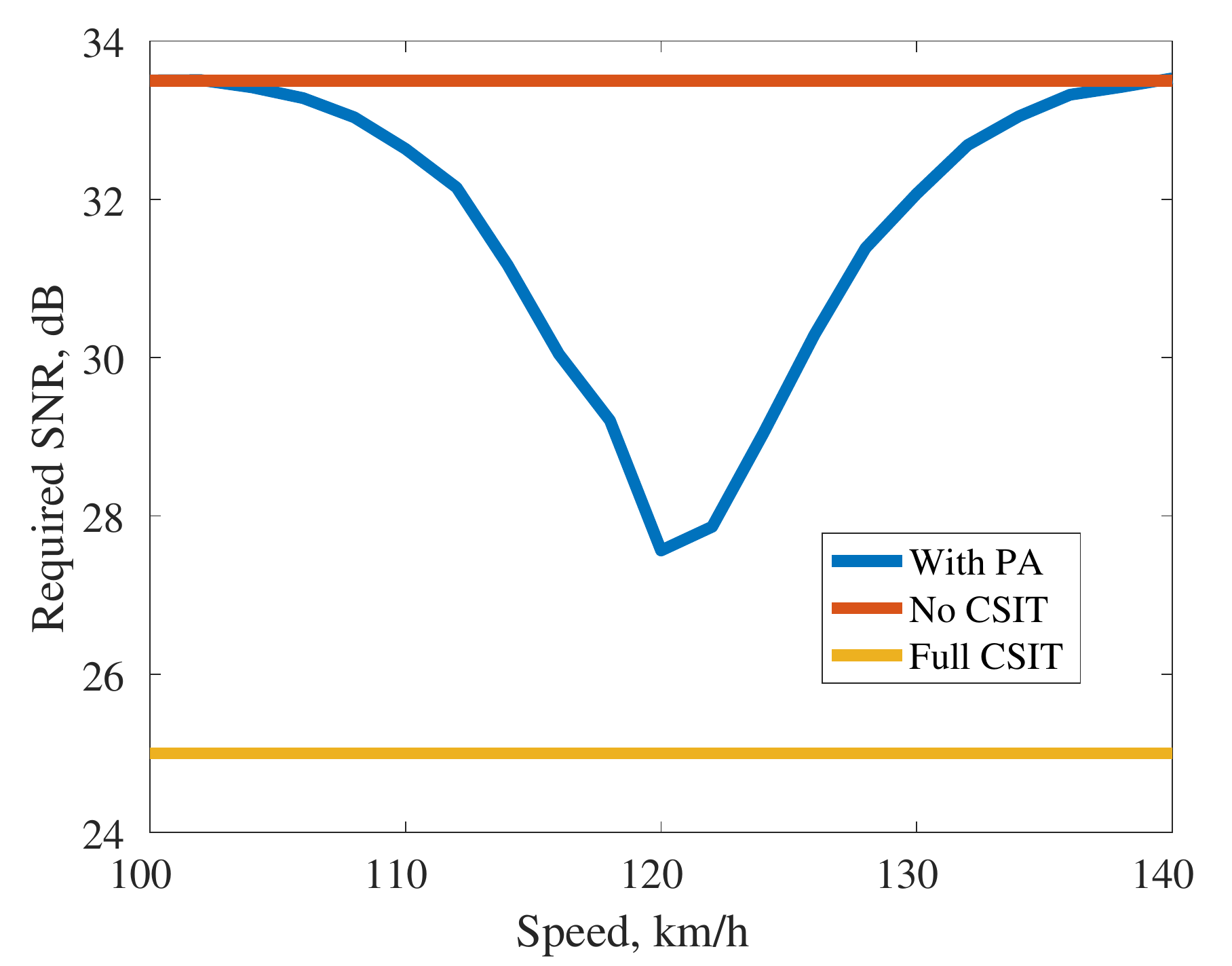}\\
\caption{Required transmit \ac{SNR} as a function of vehicle speed. Here, the required throughput is set to 5 \ac{npcu}.}
\label{fig_2}
\end{figure}

\begin{figure}
\centering
  \includegraphics[width=1.0\columnwidth]{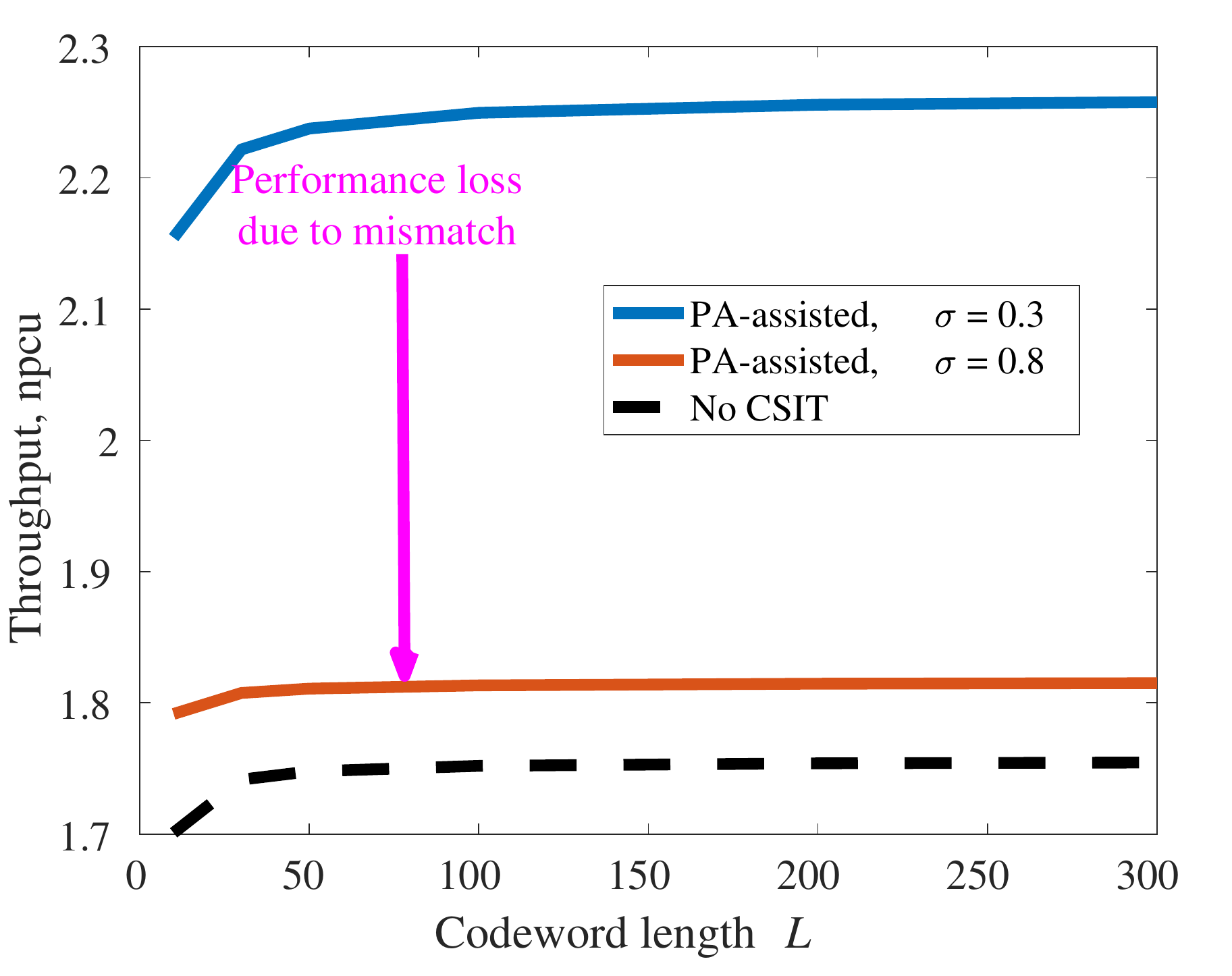}\\
\caption{Throughput as a function of the codeword length considering \ac{FBL}. $\sigma$ is a function of the mismatch distance $d$ and larger $\sigma$ represents higher mismatch. }
\label{fig_3}
\end{figure}

\section{Simulation Results}
\begin{figure}
\centering
  \includegraphics[width=1.0\columnwidth]{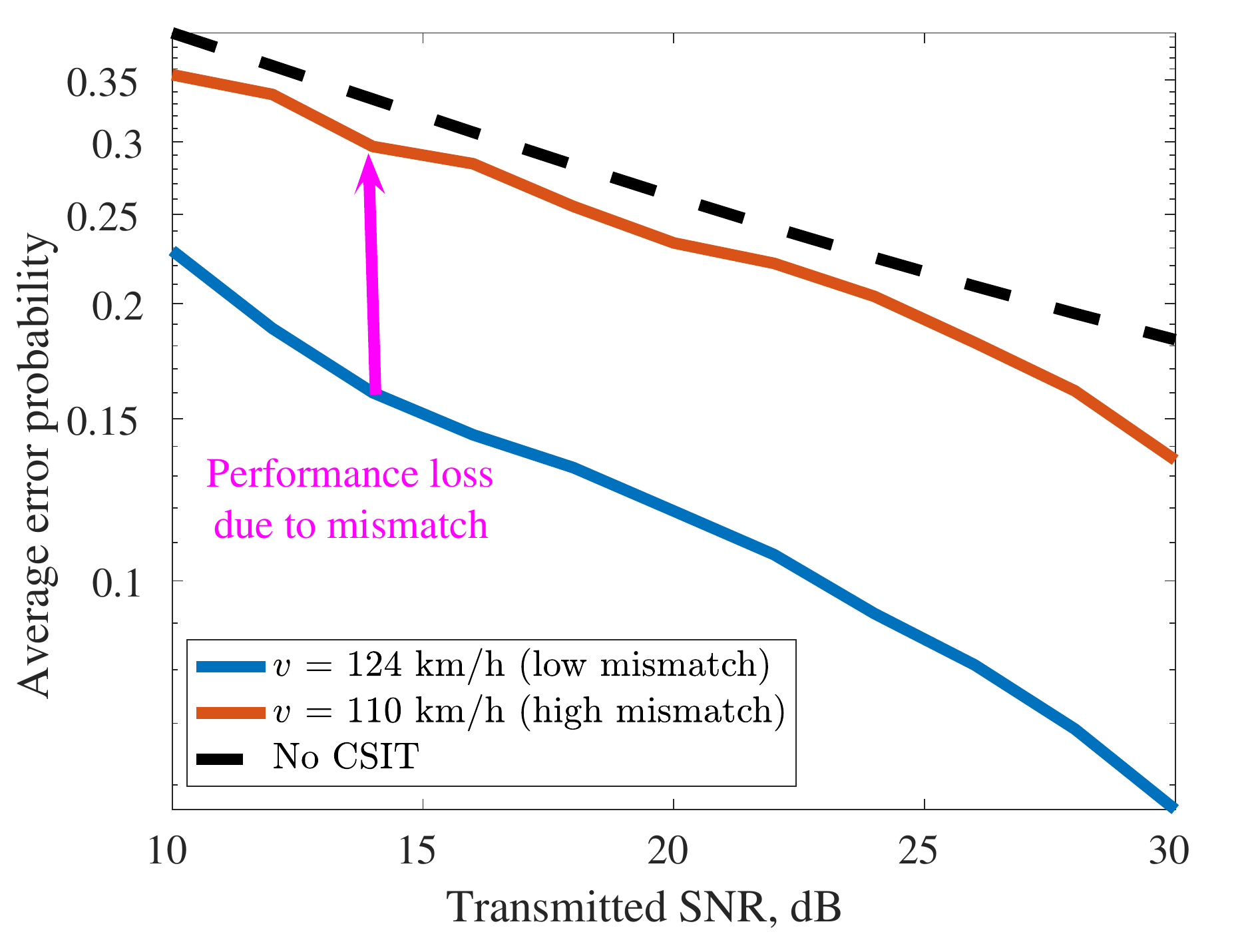}\\
\caption{Average error probability as a function of transmit \ac{SNR}. Codeword length is set to 300. According to (\ref{eq_d}), $v$ = 124 km/h is one case with low mismatch while $v$ = 110 km/h has higher mismatch. }
\label{fig_4}
\end{figure}

\begin{figure}
\centering
  \includegraphics[width=1.0\columnwidth]{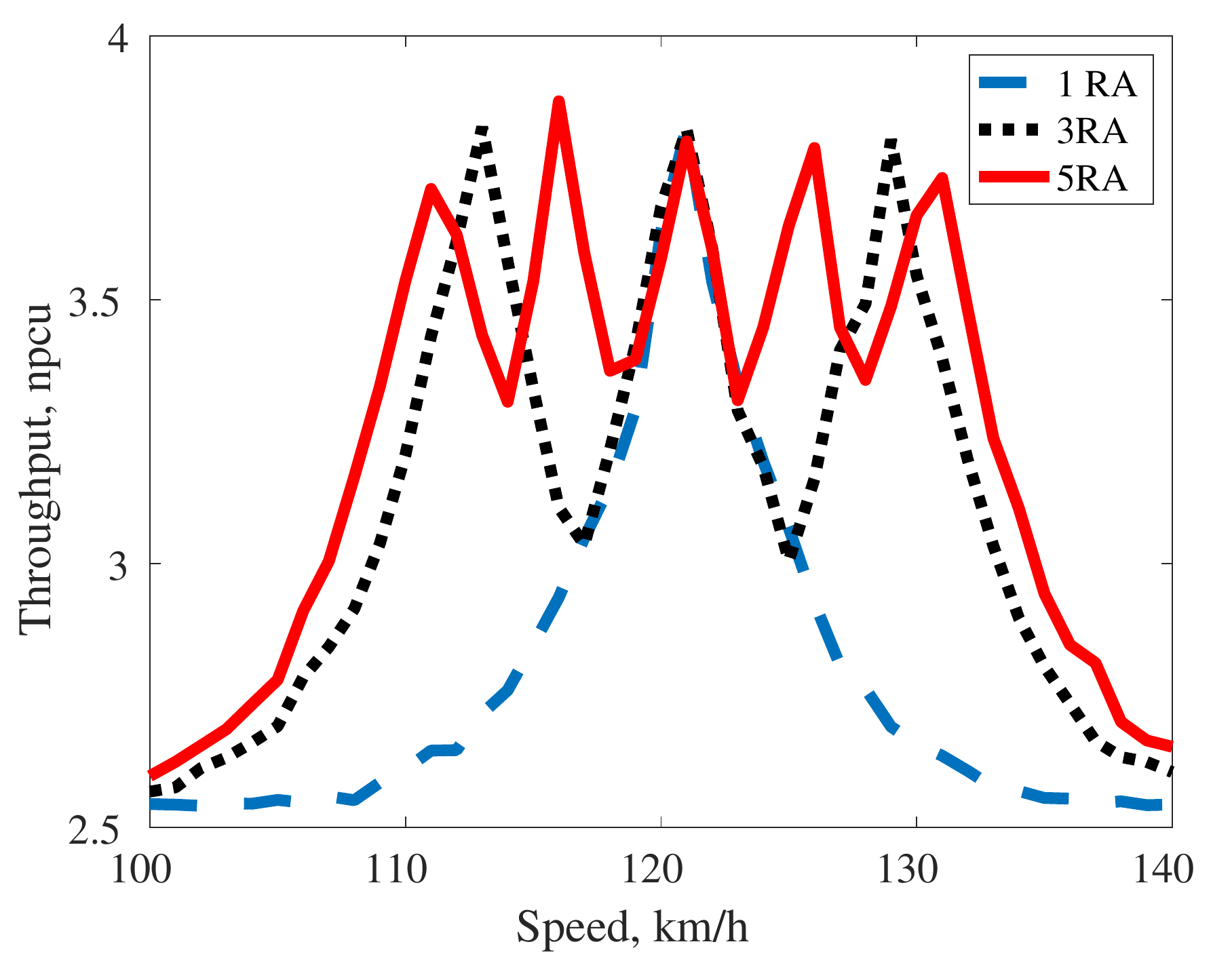}\\
\caption{Throughput as a function of the number of \acp{RA}. Algorithm \ref{alg_1} is used for optimal antenna selection. Here, the transmit \ac{SNR} is set to 20 dB. In the case with one \ac{RA}, $d_\text{a} = 1.5 \lambda$, and for three \acp{RA} $d_\text{a, 1}$, $d_\text{a, 2}$, and $d_\text{a, 3}$ are set to 1.6, 1.5 and 1.4 $\lambda$, respectively. For five \acp{RA} the antenna separations are 1.62, 1.56, 1.5, 1.44, and 1.38 $\lambda$.}
\label{fig_5}
\end{figure}

\begin{figure}
\centering
  \includegraphics[width=1.0\columnwidth]{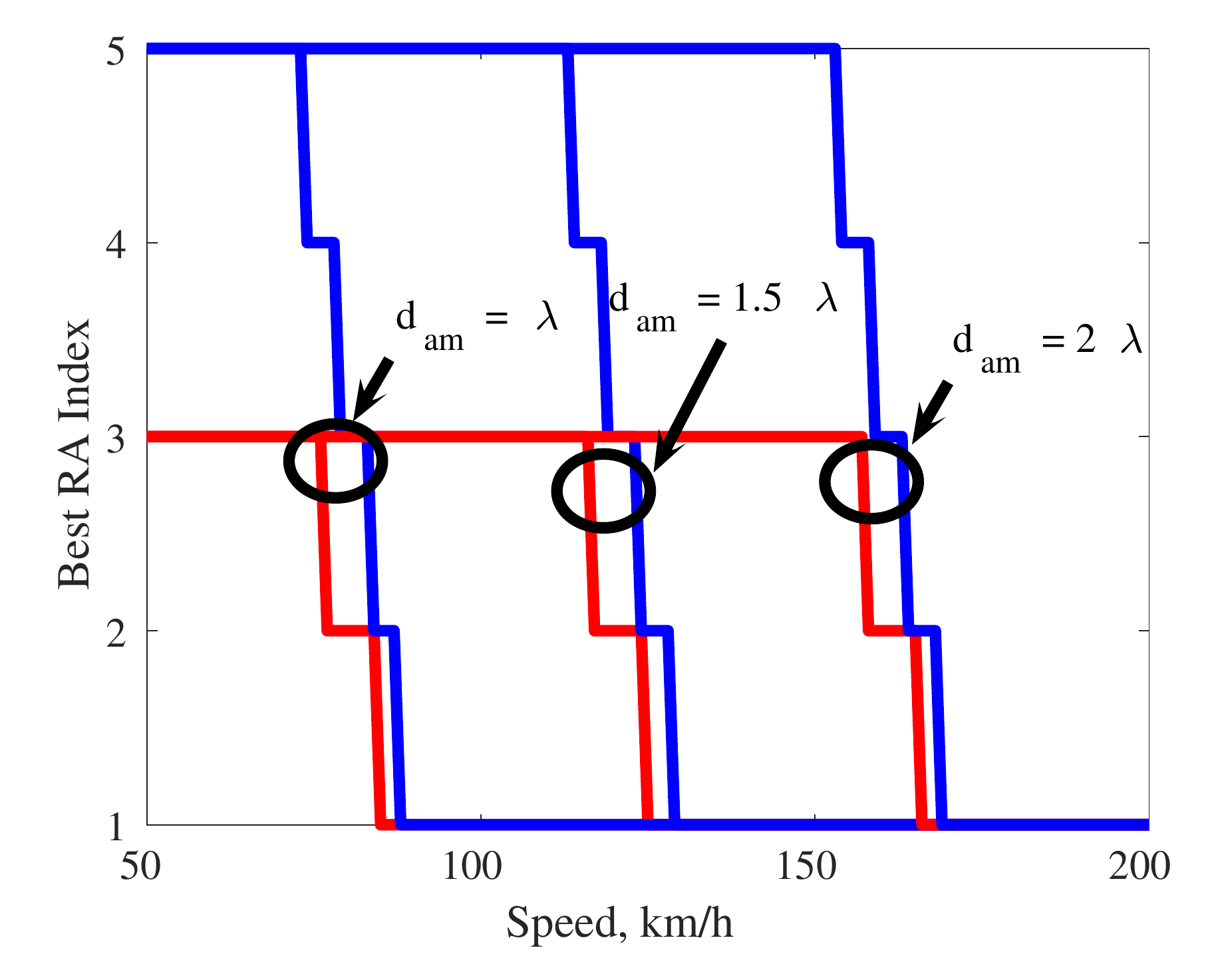}\\
\caption{The best \ac{RA} index using Algorithm \ref{alg_1}  for different values of $d_\text{am}$, which is the distance between the \ac{PA} and the \ac{RA} in the middle. The parameter settings are the same as in Fig. \ref{fig_5}.}
\label{fig_6}
\end{figure}

\begin{figure}
\centering
  \includegraphics[width=1.0\columnwidth]{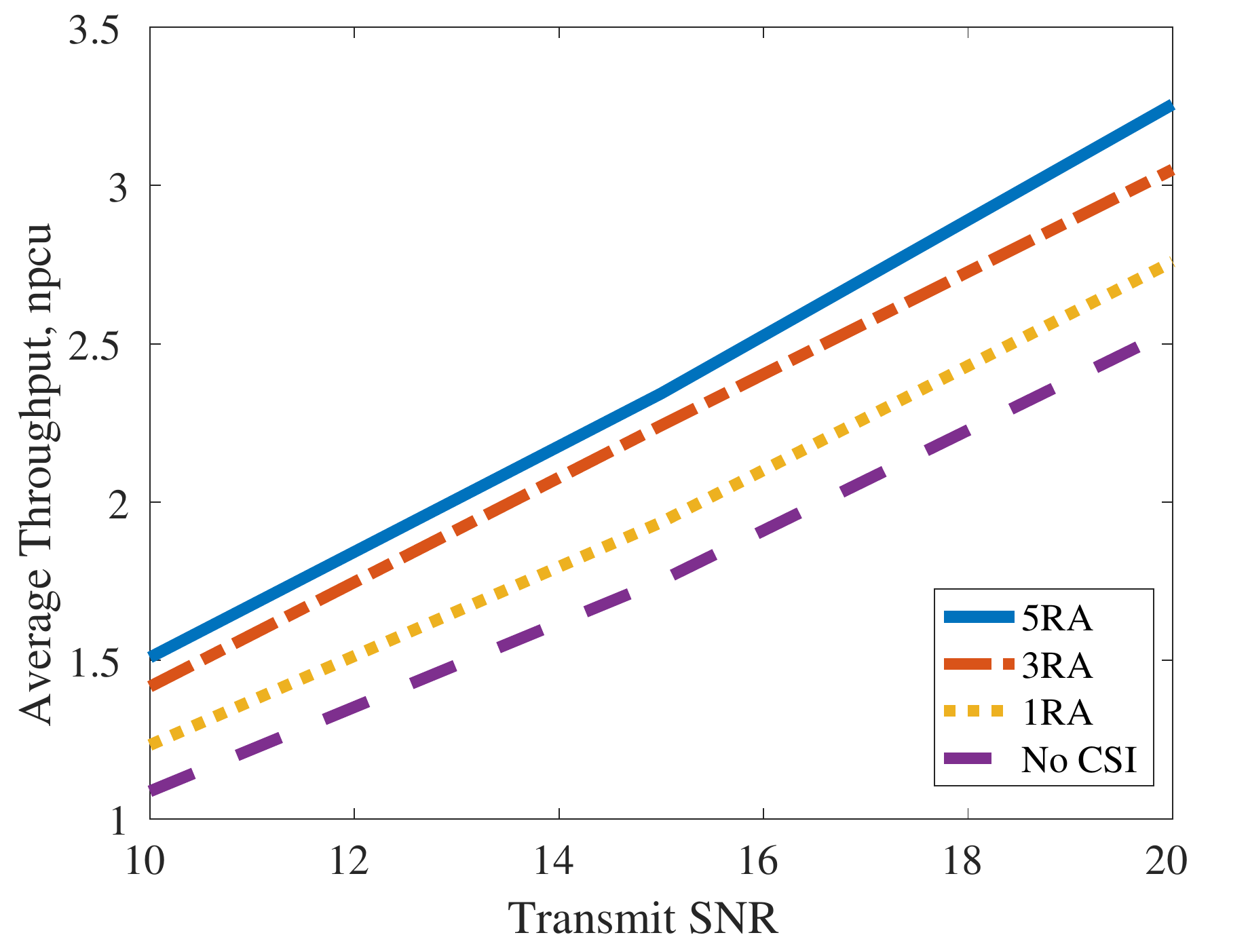}\\
\caption{Average throughput of the considered speed range (100-140 km/h) for different cases. The parameter settings are the same as in Fig. \ref{fig_5}.}
\label{fig_7}
\end{figure}

In the simulations, we study the performance of the \ac{PA} system with a focus on the effect of speed variation on the system performance. We set moving time $T$ = 5 ms and carrier frequency 2.68 GHz.  In Figs. \ref{fig_2}-\ref{fig_4} we study the performance loss of one \ac{RA} system due to spatial/speed mismatch in terms of required \ac{SNR} for given average throughput, throughput with \ac{FBL} codewords \cite{guo2020comnet}, and average error probability under \ac{FBL} \cite{guo2020comnet}, respectively, with antenna separation set to 1.5 times the wavelength. Then, in Figs. \ref{fig_5}-\ref{fig_7}, the detailed performance of the proposed antenna selection scheme is presented. 

Fig. \ref{fig_2} shows the required transmit \ac{SNR} for given throughput threshold as a function of different speed. Here, the throughput threshold is set to 5 \ac{npcu}. The case with \ac{PA} and rate adaptation is compared to the performance with full \ac{CSIT} and no \ac{CSIT}. Also, in Fig. \ref{fig_3}, considering \ac{FBL} codewords \cite{guo2020comnet}, throughput is calculated for both cases with/without the \ac{PA} for different spatial mismatches. Here, $\sigma$ is a function of the mismatch distance $d$ and larger $\sigma$ represents higher mismatch. Similarly, focusing on average error probability, Fig. \ref{fig_4} shows how it changes with transmit SNR with codeword length set to 300.

Then, applying Algorithm \ref{alg_1}, Fig. \ref{fig_5} studies the throughput as a function of vehicle speed for one, three, and five \ac{RA}(s), respectively. In the case with one \ac{RA}, $d_\text{a} = 1.5 \lambda$, and for three \acp{RA} $d_\text{a, 1}$, $d_\text{a, 2}$, and $d_\text{a, 3}$ are set to 1.6, 1.5 and 1.4 $\lambda$, respectively. For five \acp{RA} the antenna separations are 1.62, 1.56, 1.5, 1.44, and 1.38 $\lambda$. Using similar settings, i.e., the distance between \acp{RA} are the same for three and five \acp{RA}, Fig. \ref{fig_6} indicates the best \ac{RA} index for different $d_\text{am}$, which is the distance between the \ac{PA} and the \ac{RA} in the middle. Finally, in Fig. \ref{fig_7}, the average throughput for the considered speed range (100-140 km/h) are plotted for the cases with different number of \acp{RA} and no \ac{CSIT}. The antenna separations are the same as in Figs. \ref{fig_5}-\ref{fig_6}.

According to the simulation figures, the following conclusions can be drawn:
\begin{itemize}
    \item With spatial mismatch, considerable performance loss can be observed. For instance, the required \ac{SNR} increases with larger mismatch and eventually becomes equal to the case without \ac{CSIT} (Fig. \ref{fig_2}). 
    \item As shown in Fig. \ref{fig_3}, the \ac{PA} system is robust to short/moderate length packages. However, both throughput and average error probability in Fig. \ref{fig_4} are drastically affected by the speed/spatial mismatch. 
    \item Using Algorithm \ref{alg_1} and perform antenna selection based on vehicle speed, the throughput becomes much smoother with increasing number of \acp{RA}  (Fig. \ref{fig_5}). As can be seen in Fig. \ref{fig_6}, the system can be easily extended to wider speed ranges if the position of the \ac{RA} in the middle is shifted.
    \item Finally, the average throughput increases with number of \acp{RA} and using Algorithm \ref{alg_1}.
\end{itemize}

\section{Conclusions}
We reviewed the PA concept and recent testbed- and theoretical- oriented studies on the PA. With promising progress on the PA studies, it has emerges as one promising enabler for MR and moving IAB nodes in next generation of mobile networks. Dealing with the speed/spatial mismatch problem, we proposed a novel antenna selection scheme to fully utilize the velocity information and make the system more robust to speed variations. The simulation results further clarified the effect of mismatch from different perspectives, and our proposed velocity-aware antenna selection  scheme shows notable performance gains with multiple RA deployed at the vehicle side.

\section*{Acknowledgement}
This work was supported in part by VINNOVA (Swedish Government Agency for Innovation Systems) within the VINN Excellence Center ChaseOn and in part by the European Commission through the H2020 project Hexa-X (Grant Agreement no. 101015956).

\bibliographystyle{IEEEtran}
%\vspace{-2mm}
\bibliography{main.bib}

\end{document}